\documentclass[conference,compsoc]{IEEEtran}
\IEEEoverridecommandlockouts
\usepackage{cite}
\usepackage{amsmath,amssymb,amsfonts}
\usepackage{algorithmic}
\usepackage{graphicx}
\usepackage{textcomp}
\usepackage{xcolor}
\def\BibTeX{{\rm B\kern-.05em{\sc i\kern-.025em b}\kern-.08em
    T\kern-.1667em\lower.7ex\hbox{E}\kern-.125emX}}
    
\usepackage{tikz}
\usepackage{amsmath}
\usetikzlibrary{shapes,arrows, arrows.meta,matrix, positioning, fit, calc, decorations.pathreplacing}
\usepackage{booktabs}

\usepackage[inline]{enumitem}

\begin{document}

\date{}

\title{\Large \bf ``We are a startup to the core'':\\
  A qualitative interview study on the security and privacy development practices in Turkish software startups}

\author{\IEEEauthorblockN{Dilara Keküllüoğlu}
\IEEEauthorblockA{\textit{University of Edinburgh}, UK \\
d.kekulluoglu@ed.ac.uk}
\and
\IEEEauthorblockN{Yasemin Acar}
\IEEEauthorblockA{ \textit{The George Washington University}, USA \& \\
\textit{Paderborn University}, Germany \\
acar@gwu.edu}
}
\maketitle

\begin{abstract}
Security and privacy are often neglected in software development, and rarely a priority for developers. This insight is commonly based on research conducted by researchers and on developer populations living and working in the United States, Europe, and the United Kingdom. However, the production of software is global, and crucial populations in important technology hubs are not adequately studied. 
The software startup scene in Turkey is impactful, and comprehension, knowledge, and mitigations related to software security and privacy remain understudied.
To close this research gap, we conducted a semi-structured interview study with 16 developers working in Turkish software startups. The goal of the interview study was to analyze if and how developers ensure that their software is secure and preserves user privacy. 
Our main finding is that developers rarely prioritize security and privacy, due to a lack of awareness, skills, and resources. We find that regulations can make a positive impact on security and privacy. Based on the study, we issue recommendations for industry, individual developers, research, educators, and regulators. Our recommendations can inform a more globalized approach to security and privacy in software development. 
\end{abstract}


\section{Introduction}
Security and privacy research has recently focused on developers~\cite{tahaei2019survey} and their software development practices. However, most of this research has been west-centric, and has often included large, established companies, such as Microsoft~\cite{howard2006security}. It is possible that many processes are influenced by culture and/or proximity to the US both in collaboration and geography, as well as education. Company size and maturity may also play an important role for security and privacy~\cite{haney2018we}. 

A substantial share of software that users use world-wide -- both Western and non-Western users -- is created outside of the West. However, there is minimal security and privacy research to reflect this share. There are calls for more inclusive security and privacy research in the field since the systems implemented to provide security and privacy leave out the consideration of under-studied populations~\cite{wang2018inclusive}. One of these under-studied populations is Turkey which contains one of the important hubs for software development. According to Github reports~\cite{octoverse}, Turkey is one of the top 10 countries with the largest growth in software developers. The popularity of computer engineering and science degrees is also increasing in the country every passing year, even overtaking the ever popular medicine degrees.
The tech sector is also growing along with this interest in the computer science as a field. Some of the reasons why these degrees are popular in Turkey is the increasing demand for software developers and the opportunity to find jobs abroad easily, especially now that working remotely is the norm.

Alongside with the growing talent pool, the startup ecosystem in Turkey provides access to Europe, Asia, and the Middle East which is desirable for entrepreneurs~\cite{startupblink}. Hence, the startup culture in Turkey is vibrant and produces internationally used software (e.g. Getir, Udemy, Peak). However, the usability, security, and privacy mindsets and practices of developers working in these startups have not been researched. To close this gap, we conduct a 16-participant semi-structured interview study with those working in Turkish software startups, where we investigate their development processes and whether and how they relate to security, privacy, and usability. 

In this study, we answer the following research questions for Turkish startups: 
\begin{itemize}
    \item[RQ1] How do software developers define usability, security, and privacy? 
    \item[RQ2] How are usability, security, and privacy integrated into the development cycle?
    \item[RQ3] What are the possible risks developers, companies, and users face regarding security and privacy? What steps are the developers and companies taking to mitigate these risks?
\end{itemize}

We also investigate how the development process works in startups, the major important factors for developers while creating software, and the division of responsibility for the usability, security, and privacy risks discussed.

With this paper, we make the following contributions: 
We give insight into an untapped but popular and influential software scene. We focus on the developers in startups, which have different characteristics compared to small and mid-size enterprises (SMEs) or corporate companies.
We find similar relationships to security and privacy that have been published on developers that have previously been studied: we identify customer, resource and government impacts on security and privacy~\cite{tahaei2019survey} that can vary by societal and political context. 

We find that usability is easy to define and important to developers but not linked to security and privacy; security and privacy are harder to define and seem less achievable early in the startup. Developers see security and privacy as intertwined, and consider security as a means to achieve privacy. 

Based on these insights, we make recommendations for developers, researchers and educators, regulatory bodies, as well as for developing software industries to improve software security and privacy.

\section{Related Work}

After giving a brief background on Turkey, we discuss related work in three main areas: human factors research with software developers for security and privacy, research on developer processes and behavior in startups, as well as research on software developers in Turkey. 

\subsection{Background on Turkey}

Turkey has a personal data protection law, Ki\c{s}isel Verilerin Korunmas\i{} Kanunu (KVKK)~\cite{kvkk}, which came into effect in 2016. Every organization operating in Turkey or processing data of Turkish residents is required to abide by this mandate.  

While computer science has become one of the most popular degrees, as mentioned in Introduction, the representation of women in companies as software developers stays comparatively low. The percentage of women working as information and communication technology (ICT) specialist has increased minimally from 2011 to 2020 from 14.3\% to 16.8\%~\cite{eurostat}. Even when women work as ICT specialists, they are pushed to roles that are perceived as more ``feminine'' such as analysts instead of developers~\cite{bozkurt2017bilicsim}.

Turkey is a European Union candidate while also regarded as a non-WEIRD country by prior research~\cite{muthukrishna2020beyond, klein2018many}. We acknowledge that Turkey cannot be representative of all non-WEIRD countries, which include many countries from a wide range of regions around the world. Our study adds a facet to the body of literature on diverse groups of software developers.

\subsection{Security \& Privacy Research with Developers}

While there is an extensive research on usable security and privacy for end users, researchers recently started focusing on developers and their mindsets around security and privacy. Acar et al.~\cite{acar2016you} provided a research roadmap to study developers in the context of usable security and privacy. Green and Smith~\cite{green2016developers} also called for developer friendly security and assisting the developers in various ways such as more usable security APIs and libraries, better tools for security testing, as well as safer programming languages. 

Assal and Chiasson~\cite{assal2018security} interviewed 13 participants to understand the consideration of security in development pipelines as well as how this process is affected by the security knowledge of the software developers. They found that division of workload, expectation of security knowledge from developers, company culture, and availability of resources are some of the factors affecting the security practices in companies. External pressure from entities like customers and government mandates also influences the adoption of security in their development process. Assal and Chiasson~\cite{assal2019think} followed up on this study with a quantitative survey study to further explore the software security practices and developers' role in these. They found that, rather than developers' reluctance, a lack of support from the companies prevents the creation of secure products. Conversely, Bednar et al.~\cite{bednar2019engineering} found that developers were not motivated to implement privacy-by-design principles citing lack of social pressure, clarity, and technical difficulty.

Tahaei et al.~\cite{tahaei2021privacy} conducted interviews with developers who give extra importance to advocating end-user privacy in companies. They called these developers ``Privacy Champions'' and found that these developers can successfully improve the privacy culture of their organizations if they are supported by the company and their peers.

Hadar et al.~\cite{hadar2018privacy} conducted interviews with 27 developers from Israel and India to understand their perceptions and practices of protection of data. They find that developers refer to data security when asked about privacy resulting in limiting privacy to third-party attacks. Developers do not have required knowledge to implement privacy preserving technologies and the companies have a great effect on developers' privacy behavior. Poller et al.~\cite{poller2017can} also show that security training given to developers is not enough by itself, companies should also provide a suitable environment for developers to implement what they have learned. Bu et al.~\cite{bu2020privacy} add that companies should also give incentives to developers to employ best practices.

\subsection{Research on Developers in Startups}

Most of the studies mentioned were focusing on SME and Large Enterprises where our study is on startups. Compared to these companies, startups differ in various ways including lack of resources (e.g. human, funds, time), higher uncertainty levels, as well as higher risk~\cite{giardino2014we}. They are also highly innovative and reactive to changes. 

Souza et al.~\cite{souza2017software} studied four software startups founded in an academic environment to understand their development practices. To do so, they conducted a case study that included various data including semi-structured interviews and questionnaires. They find that implementing standard development practices is not trivial for early stage startups. Lack of resources and uncertainty about the fitness of the product move startups to adopt a more evolutionary approach in their product development. In a more recent work, Souza et al.~\cite{souza2021survey} conducted a survey with 140 developers who work in Brazilian startups to again study software engineering practices but this time from the developers' perspective. They find that the skills and knowledge of the development team drive the software architecture decisions. Their participants also considered usability, performance, and security as important quality attributes to focus on.

Zhang and Xie~\cite{zhang2018toward} conducted a diary study along with in-depth interviews with IoT developers in China to understand their development features and patterns. They found that IoT developers often take on more than one role in the startups, third-party open source projects are frequently used, and lack of usable developing tools is a major pain point. They also found that security issues were not prioritized by their participants compared to global IoT developers.

While these studies focus on startup developers, security and privacy are only mentioned in passing, with minimal studies solely focusing on it~\cite{paternoster2014software}. Balebako et al.~\cite{balebako2014privacy} studied the privacy and security behaviours of smartphone app developers in the United States, mostly with participants from companies with 2-9 employees. They found that many participants did not receive formal training on privacy and security, and compared to bigger companies, smaller companies were less conscious about privacy.

\subsection{Developer research in Turkey}

Software practices of developers and companies in Turkey remain understudied, especially when it comes to startups, security, and privacy. The existing research is often published in Turkish, which is not easily accessible to the rest of the world, and may be prohibitive for researchers to compare and contrast with their research.

Garosi et al.~\cite{garousi2015survey} surveyed 202 participants that were working in companies that have development offices in Turkey to understand software engineering practices in Turkey. They found that rather than test-driven development, companies still use test-last development prominently. While functional and system testings are common, they use stress testing and security testing least widely. In this work, there were limited mentions of security and no focus on privacy and the companies were in mixed sizes with 33\% of them having more than 500 employees. In our work, we focus on privacy and security perceptions of developers working in startups.

Y{\i}lmaz~\cite{yilmaz2019observed} observed the process of changing software practices for three companies in Turkey to get insights into the process, the results, and find the factors affecting the adoption of new practices. They conducted semi-structured interviews with 18 participants and found that software developers, software testers, and the team leaders have the most impact on the process. Many of the participants associated the changing process with requirements analysis; security and usability is cited as important factors affecting the development process along with other factors such as cost and customer satisfaction.

Studies focusing on software developers in Turkey are minimal. Those that do study software development in Turkey do not specifically explore startups, security, or privacy. In our work, we are investigating the security and privacy perceptions and processes of startup developers in Turkey to fill the gap in the literature.

\subsection{Regulations by Organizations}

As Assal and Chiasson~\cite{assal2018security} found, external pressures from government mandates and policies drive companies to adopt security practices. The European Union (EU) implemented the General Data Protection Regulation (GDPR) in 2016 and enforced it in 2018. GDPR regulates data protection, privacy, as well as personal data handling in the EU countries. Similarly, the State of California enforced the California Consumer Privacy Act (CCPA)~\cite{ccpa}, which deals with consumer data privacy in 2020. Turkey also has its own personal privacy protection regulation called Ki\c{s}isel Verilerin Korunmas\i{} Kanunu (KVKK)~\cite{kvkk}, translated as Personal Data Protection Law, which became effective earlier than both GDPR and CCPA. KVKK regulates the handling of user data by companies, and came to effect in 2016. However, prior research around GDPR shows that these regulations are difficult to implement for developers because of the lack of familiarity and guidance~\cite{alhazmi2020struggle}. 
In extension of existing work on the adoption and implementation of the GDPR, we shed light into  the adoption and practices by startup developers related to the KVKK in Turkey.

\section{Methodology}

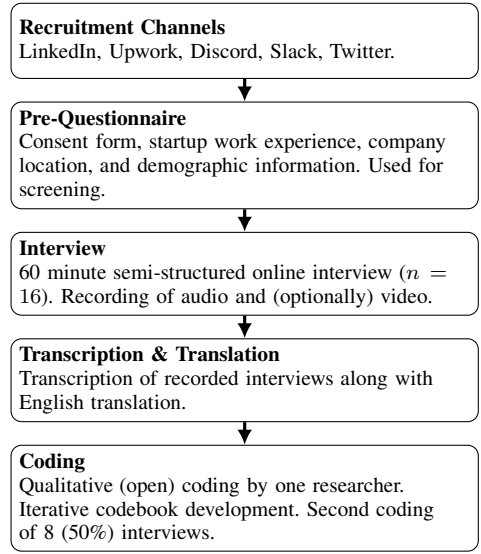
\begin{figure}[t]
	\centering
	\newcommand\interviews{16}
\begin{tikzpicture}[>={Latex[width=2mm,length=2mm]},
    node distance = 0.3cm and 0.6cm,
    block/.style = {rectangle, rounded corners, draw=black, minimum width=4.5cm, minimum height=1cm, align=left, text width=60mm, inner sep=0.3em},
    recchannel/.style = {minimum width=1.2cm},
    auto,
    font=\footnotesize
]

\node (recruitment) [block]
{\textbf{Recruitment Channels}\\LinkedIn, Upwork, Discord, Slack, Twitter.};
\node (questionnaire) [block, below =of recruitment] {\textbf{Pre-Questionnaire}\\Consent form, startup work experience, company location, and demographic information. Used for screening.};
\node (interview) [block, below =of questionnaire] {\textbf{Interview}\\60 minute semi-structured online interview ($n=\interviews$). Recording of audio and (optionally) video.};
\node (transcription) [block, below =of interview] {\textbf{Transcription \& Translation}\\Transcription of recorded interviews along with English translation.};
\node (coding) [block, below =of transcription] {\textbf{Coding}\\Qualitative (open) coding by one researcher. Iterative codebook development. Second coding of 8 (50\%) interviews.};

\draw[->, very thick] (recruitment) -- (questionnaire);
\draw[->, very thick] (questionnaire) -- (interview);
\draw[->, very thick] (interview) -- (transcription);
\draw[->, very thick] (transcription) -- (coding);

\end{tikzpicture}
	\caption{Methodology overview.}
	\label{fig:overviewmethodology}
\end{figure}

We conducted a semi-structured interview study of 16 people who work in Turkish startups as software developers or founded them between 17th Aug - 2nd November 2021. For readability, we call all participants ``developers'' from here on, specifying if they are a founder when it relates to the answer they give. For this work, startups are defined as companies who were founded in the last five years (i.e. from 2016) and are in the technology sector. We describe our methodology in this section (see Figure~\ref{fig:overviewmethodology}).

\subsection{Positionality Statement}
Positionality statements help to give more context into the study and inform the reader on how the researchers' culture and experiences shaped the research~\cite{schlesinger2017intersectional,liang2021embracing}. Dilara Keküllüoğlu, the first author of this paper and the sole interviewer, is fluent in Turkish and has worked in the Turkish software startup scene in the past. The increasing number of the startups originating from Turkey, demand for software developers in the country, as well as the lack of published research focusing on developers in Turkey motivated this paper. Her experiences helped to inform which questions to ask, facilitated recruitment of participants, and enabled her to conduct interviews in Turkish, the native language of the participants; her work experience within the Turkish startup scene helped connect with participants and contextualize findings. The interviewer is a woman in a similar age bracket with the participants, which helped establish rapport with interviewees, and she is the only author with Turkish language fluency.

\subsection{Participant Recruitment}

For this study we wanted to find: software developers who (1) are Turkish or lived in Turkey more than 10 years, (2) worked in technology startups in Turkey that were founded in the last five calendar years, (3) are able to have an interview in Turkish or English. To do so, we created a prescreening survey on Qualtrics and distributed the link over social media channels such as LinkedIn, Twitter, and Discord, and Slack channels relating to the Turkish startup scene. 
We also recruited participants from Upwork where we prescreened them using the internal screening function. We recruited 16 participants in total where 4 were from social media channels and 12 were from Upwork.

After checking the eligibility of the participants, we asked additional questions including ones about their company, working experiences, and demographic information. At the end of the survey, they were directed to an online scheduling service to arrange a meeting time (\textasciitilde60mins). 

\subsection{Demographics}

Out of 16 participants, 12 (75\%) were men and four (25\%) were women. The average age of participants is 28.4 (median 25.5, sd. 5.9). Nearly all of the participants (15, 94\%) obtained a bachelor's degree while the remaining person completed an associate's degree. Many of the participants (6, 38\%) have computer science (computer engineering) degrees. The most experienced participant has 23 years of software development experience while on average our participants have 6.9 years of experience (sd. 6.3). We share the detailed demographics in Table~\ref{tab:demographics}.

\newcommand\interviewsDemoGenderMale{12}
\newcommand\interviewsDemoGenderMalePerc{75\%}
\newcommand\interviewsDemoGenderFemale{4}
\newcommand\interviewsDemoGenderFemalePerc{25\%}

\newcommand\interviewsDemoAgeMin{22}
\newcommand\interviewsDemoAgeMax{43}
\newcommand\interviewsDemoAgeMean{28.4}
\newcommand\interviewsDemoAgeStd{5.9}
\newcommand\interviewsDemoAgeMedian{25.5}

\newcommand\interviewsDemoExpMin{1}
\newcommand\interviewsDemoExpMax{23}
\newcommand\interviewsDemoExpMean{6.9}
\newcommand\interviewsDemoExpStd{6.3}
\newcommand\interviewsDemoExpMedian{5}

\newcommand\interviewsDemoEduBachelor{15}
\newcommand\interviewsDemoEduBachelorPerc{94\%}
\newcommand\interviewsDemoEduTwoYear{1}
\newcommand\interviewsDemoEduTwoYearPerc{6\%}

\newcommand\interviewsDemoPositionSWE{8}
\newcommand\interviewsDemoPositionDataSci{2}
\newcommand\interviewsDemoPositionFounder{3}
\newcommand\interviewsDemoPositionFrontend{2}
\newcommand\interviewsDemoPositionCTO{1}
\newcommand\interviewsDemoPositionArch{1}
\newcommand\interviewsDemoPositionTeamLead{1}

\begin{table}[t]
    \caption{Participants' demographics. Two of the co-founders are also working as developers.}
    \label{tab:demographics}
    \centering
    \footnotesize
    
    \setlength{\tabcolsep}{0.66\tabcolsep}
    \setlength{\defaultaddspace}{0.25\defaultaddspace} 

    \begin{tabular}{llrrlrr}
        \toprule
        \multicolumn{7}{l}{\textbf{Gender}}  \\
        & Man & \interviewsDemoGenderMale & (\interviewsDemoGenderMalePerc) \\
        & Woman & \interviewsDemoGenderFemale & (\interviewsDemoGenderFemalePerc)\\
        
        \addlinespace
        \multicolumn{7}{l}{\textbf{Age [years]}} \\
        & Min. & \interviewsDemoAgeMin & & Max. & \interviewsDemoAgeMax & \\
        & Mean (Std.) & \interviewsDemoAgeMean{} & $\pm$\interviewsDemoAgeStd & Median & \interviewsDemoAgeMedian \\
        \addlinespace
        \multicolumn{7}{l}{\textbf{Industry Experience [years]}}\\ 
        & Min. & \interviewsDemoExpMin & & Max. & \interviewsDemoExpMax & \\
        & Mean (Std.) & \interviewsDemoExpMean{} & $\pm$\interviewsDemoExpStd & Median & \interviewsDemoExpMedian \\
        \addlinespace
        \multicolumn{7}{l}{\textbf{Education}}  \\
         & Bachelor's degree                      & \interviewsDemoEduBachelor & (\interviewsDemoEduBachelorPerc)  & \\
         & Some college or \\ & two-year associate degree          & \interviewsDemoEduTwoYear & (\interviewsDemoEduTwoYearPerc) \\
         \addlinespace
        \multicolumn{7}{l}{\textbf{Position in Company}}  \\
         & Software Developer & \interviewsDemoPositionSWE & 
         & Co-founder* & \interviewsDemoPositionFounder & \\
         & Data Scientist & \interviewsDemoPositionDataSci &
         & Front-end Developer & \interviewsDemoPositionFrontend & \\
         & CTO & \interviewsDemoPositionCTO &
         & Solutions Architect & \interviewsDemoPositionArch & \\
         & SWE Team Lead & \interviewsDemoPositionTeamLead & \\
         \bottomrule
    \end{tabular}
\end{table}

\subsection{Ethics \& Data Protection}

The prescreens started with a participant information sheet in Turkish where we described the study, data collection and storage, their rights regarding the study, and gave contact information. Only consenting participants could continue to next questions in the survey. In the interview, we once again confirmed that participants were consenting to have their voices recorded and transcribed by GDPR-compliant services before starting the recording. Participants were reminded of their right to skip any question they wished and we answered their questions regarding the study and data storage concerns. We paid our participants the equivalent of \$17-20 (fluctuations were due to currency changes and varying recruitment channel overheads) which is above the average hourly rate of software developers in Turkey.

We used a secure cloud to keep recordings, survey results, and transcriptions. The transcription and translation service we used is GDPR-compliant and the transcripts were de-identified before starting coding on them. Identifying information (like payment information) was processed immediately, and stored separately from study data. Our study was approved by our institution's ethical review board. 

\subsection{Interview Procedure}
After prescreening for participants who work in startups in Turkey, we asked them questions about their company, their role in the company, software developing experience as well as some demographic questions such as age and gender. We also gave our participants the choice to hold the interview in Turkish or English. The screening questionnaire can be found in the Appendix~\ref{app:prescreen}.
We conducted semi-structured interviews with four main sections: (1) background and education, (2) current position and the project, (3) development life-cycle, and (4) usability, security, and privacy in software development.
While we were mainly interested in security and privacy in the development context; we also talked with developers about usability, as relates both to security and privacy~\cite{garfinkel2014usable}, and may be neglected in favor of primary functionality goals, similar to security and privacy.
Semi-structured interviews gave us in-depth answers from our participants about their mental models and perceptions around usability, security, and privacy and enabled us to engage with and follow up on participants' answers. We developed a first draft of our interview guide based on prior work and the experiences of the researcher who worked in Turkish startups. The interviewer conducted one pilot interview in English and another one in Turkish, after which we discussed and made minor adjustments to the interview guide. We followed the interview protocol outlined by Rader et al.~\cite{rader2020have} to conduct our interviews. Fifteen interviews were conducted in Turkish; one was in English. We did not mention usability, security, or privacy before the dedicated part in the interview guide, and allowed participants to bring them up organically. We debriefed the interviews regularly with the research team, and decided that we reached saturation with the 15th interview after the following criteria applied: (1) we received no different answers to the interview questions targeted at security and privacy; (2) no new themes, strategies, or insights emerged~\cite{saunders2018saturation,corbin2014basics}. We followed up with an additional interview after reaching saturation. The detailed interview guide can be found at Appendix~\ref{app:guide}. Interview recordings ranged from 35 to 76 minutes, with an average of 53 minutes.

\subsection{Interview Data Analysis}

For joint data analysis, the interviews were transcribed and translated to English manually by an external GDPR-compliant service, and checked for correctness by the first author, who is fluent in Turkish. The initial draft of the codebook was created according to the semi-structured interview questions. Using this codebook, the first author coded two of the interviews at first. We then discussed the coding and updated the codebook accordingly. The first author then coded the remaining interviews, and we jointly discussed the findings. Based on these discussions and peer reviews, the codebook was adjusted accordingly. The first author updated their codes for all of the interviews while the second author coded eight (50\%) of the interviews. Our disagreements were usually cosmetic (e.g., different snippet sizes were coded with the same code by the two coders).
We resolved the disagreements over codes through discussions, which helped us refine the code definitions, and established trust that we had created a meaningful and appropriate codebook and coding process~\cite{mcdonald2019reliability}. The inter-rater agreement between coders was satisfactory~\cite{lazar2017research} with Brennan-Prediger kappa~\cite{brennan1981coefficient} of .76 (between .70 and .85) when calculated with MaxQDA software~\cite{maxqda} using 80\% text segment overlap. The first coder updated the remaining eight interviews according to the finalized codebook and coding strategy. We share our codebook in Appendix~\ref{app:codebook}. We used affinity diagramming to analyze the code segments that are most relevant to the research questions, namely segments for considerations, definitions, and problems around security, privacy, and usability.

\subsection{Limitations}

Our study is an interview study that relies on self-report of the participants which may lead participants to filter their responses and report desirable behaviors more than undesirable ones~\cite{donaldson2002understanding}. 
Our participants are relatively young and most of them were men. This is expected, considering over 75\% of the software developers in Turkey are younger than 30 and only 12\% of the developers in Turkey are women. Additionally, women developers find startups risky to work in compared to men~\cite{womendevelopers}. To increase women participation in our study, we shared our recruitment post with Slack groups dedicated to women developers in Turkey, individually invited women developers in Upwork, and asked Twitter users with big woman developer following to share our post. 

We deliberately did not give any formal definitions of security, privacy, or usability in the interviews. We prioritized investigating the developers' understanding, thoughts, experiences, and processes over teaching a formal definition and possibly interrupting the interview flow or intimidating participants. As a result, we did not discuss the exact same concepts with every participant, but rather based our interview on their understanding.
\section{Results}

\subsection{Companies}
The companies that employ our participants have been operating for an average of 2.75 years (median: 2.75, standard deviation 1.3). Nine (56\%) of the 16 companies have 1-10 employees, five (31\%) have 11-50 employees. One of the remaining two has 51-100 while the other one has 101-500 employees. Our participants' teams average 4.4 members. Out of 16 companies our participants worked in, four (25\%) of them were in Software sector, four (25\%) of them were in finance technology. The remaining half of the companies were in various fields including Data Science, E-Commerce, and Mental Health. Detailed information on the companies can be found in Table~\ref{tab:companies}.

\newcommand\interviewsCompAgeMin{1}
\newcommand\interviewsCompAgeMax{5.5}
\newcommand\interviewsCompAgeMean{2.75}
\newcommand\interviewsCompAgeStd{1.3}
\newcommand\interviewsCompAgeMedian{2.75}

\newcommand\interviewsCompSizeTen{9}
\newcommand\interviewsCompSizeTenPerc{56\%}
\newcommand\interviewsCompSizeFifty{5}
\newcommand\interviewsCompSizeFiftyPerc{31\%}
\newcommand\interviewsCompSizeHund{1}
\newcommand\interviewsCompSizeHundPerc{6\%}
\newcommand\interviewsCompSizeFiveHund{1}
\newcommand\interviewsCompSizeFiveHundPerc{6\%}

\newcommand\interviewsCompTeamMin{2}
\newcommand\interviewsCompTeamMax{11}
\newcommand\interviewsCompTeamMean{4.4}
\newcommand\interviewsCompTeamStd{2.2}
\newcommand\interviewsCompTeamMedian{4}

\newcommand\interviewsCompSectorSoftware{4}
\newcommand\interviewsCompSectorSoftwarePerc{25\%}
\newcommand\interviewsCompSectorFintech{4}
\newcommand\interviewsCompSectorFintechPerc{25\%}
\newcommand\interviewsCompSectorDataSci{1}
\newcommand\interviewsCompSectorDataSciPerc{6\%}
\newcommand\interviewsCompSectorRnD{1}
\newcommand\interviewsCompSectorRnDPerc{6\%}
\newcommand\interviewsCompSectorECommerce{1}
\newcommand\interviewsCompSectorECommercePerc{6\%}
\newcommand\interviewsCompSectorCallCenter{1}
\newcommand\interviewsCompSectorCallCenterPerc{6\%}
\newcommand\interviewsCompSectorRemote{1}
\newcommand\interviewsCompSectorRemotePerc{6\%}
\newcommand\interviewsCompSectorMental{1}
\newcommand\interviewsCompSectorMentalPerc{6\%}
\newcommand\interviewsCompSectorRetail{1}
\newcommand\interviewsCompSectorRetailPerc{6\%}
\newcommand\interviewsCompSectorInd{1}
\newcommand\interviewsCompSectorIndPerc{6\%}

\begin{table}[t]
    \caption{Companies.}
    \label{tab:companies}
    \centering
    \footnotesize
    
    \setlength{\tabcolsep}{0.66\tabcolsep}
    \setlength{\defaultaddspace}{0.25\defaultaddspace} 

    \begin{tabular}{llrrlrr}
        \toprule
        \multicolumn{7}{l}{\textbf{Company Age [years]}} \\
        & Min. & \interviewsCompAgeMin & & Max. & \interviewsCompAgeMax & \\
        & Mean (Std.) & \interviewsCompAgeMean{} & $\pm$\interviewsCompAgeStd & Median & \interviewsCompAgeMedian \\
        \addlinespace
       \multicolumn{7}{l}{\textbf{Company Size}} \\
        & 1-10      & \interviewsCompSizeTen &  (\interviewsCompSizeTenPerc) & 11-50 & \interviewsCompSizeFifty & (\interviewsCompSizeFiftyPerc) \\
        &51-100 & \interviewsCompSizeHund &  (\interviewsCompSizeHundPerc) & 101-500 & \interviewsCompSizeFiveHund & (\interviewsCompSizeFiveHundPerc) \\
        \addlinespace
        
        \multicolumn{7}{l}{\textbf{Participant Team Size}}\\ 
        & Min. & \interviewsCompTeamMin & & Max. & \interviewsCompTeamMax & \\
        & Mean (Std.) & \interviewsCompTeamMean{} & $\pm$\interviewsCompTeamStd & Median & \interviewsCompTeamMedian \\
        \addlinespace
        \multicolumn{7}{l}{\textbf{Company Sector}}  \\
         & Software                     & \interviewsCompSectorSoftware &
         & Fintech       & \interviewsCompSectorFintech \\
         & Data Science                     & \interviewsCompSectorDataSci &
         & R\&D       & \interviewsCompSectorRnD \\
         & E-Commerce                     & \interviewsCompSectorECommerce &
         & Call Center       & \interviewsCompSectorCallCenter \\
         & Remote Support SW                     & \interviewsCompSectorRemote &
         & Mental Health       & \interviewsCompSectorMental \\
         & Retail                     & \interviewsCompSectorRetail &
         & Industry \\ & & & & \& Agriculture     & \interviewsCompSectorInd \\
         
         \bottomrule
    \end{tabular}
\end{table}

\textbf{Products} Our participants worked on diverse products, ranging from creating applications for smart glasses, pathway finding for industrial navigation robots, providing optimal energy generation for windmills, and establishing infrastructure for call centers.

\textbf{User portfolio} The products were mostly targeted towards other businesses with business-to-business (B2B) models (9, 56.3\%); only with four (25\%) of the companies worked exclusively with end users. Three companies provided their services to both, other businesses and end users. One of the participants mentioned their product is a market leader in Japan. Hence, after half of the interviews, we started to explicitly ask about the location of the target users. All of the companies mainly targeted Turkish users and businesses; however, five of the eight participants we asked also provided service to non-Turkish users and companies. The remaining three participants expressed interest to expand their services to other countries in the future.

\subsection{Development Pipeline}
Understanding the development process, contextualizes the environment for usability, security, and privacy by helping us understand whether the process is set up to consider these from the start, throughout the process, when implementing existing components, and in testing.
We asked our participants to walk us through their startup's pipeline for developing software in the company. We were also interested in finding out whether their development process was systematic, or more ad-hoc. Seven participants mentioned adherence to agile frameworks like scrum in their development process. Additionally, five more participants mentioned weekly meetings or plannings, which we understand to be similar to standup-meetings.

Participants reported that features that will be implemented are decided on by various teams, including management, sales teams, team leads, and product managers. For some companies with a small number of employees, decisions are made by the development team. 

Nine companies have very detailed development processes, which may include having multiple teams involved from the decision to implement to testing, having a clear backlog to live roadmap, and testing with a small user base before going live. Most of these companies have more than ten employees. Two participants mentioned that there is no established system for development process in their company. \textit{``Unfortunately not. We are trying to save the day now. This is something that actually bothers me'' (P9)}. The remaining five of them have a development process they follow but not as detailed as the first group.

\textbf{Builder of the pipeline:} Eight (50\%) of the participants stated that the pipeline is the results of a joint effort that can include founders, team leaders, and software developers. Founders established the pipeline for four (25\%) of the companies. P1 said that he built it, since he is the CTO. One of the participants stated the head of engineering of the company set up the development pipeline with the help of engineering manager, while the one said there was no pipeline. The remaining participant did not know who was the builder of the pipeline.

\textbf{Testing:} We asked participants whether they test their software and what they are testing against if so. Nearly all participants (13) mentioned doing functional tests, usually performed by the developers themselves, as one participant illustrated: \textit{``Let's not call it a separate team, it's still within ourselves. It could be said, let's test today, it will be an hour test, we will test a feature'' (P13).} Two of these participants said they will ask another person outside the team to check after they test the software themselves. \textit{``We test the code beforehand, we test it as developers. Afterwards, we have a friend who works as an analyst but is not exactly an analyst, we have him test it'' (P9)}. Four companies have an in-house testing team while three outsource their tests. Five of the companies also either use an outsider person or customer base to test. Additionally, six participants mentioned doing security tests, with two of the companies outsourcing these tests. These security tests were conducted on e-commerce, e-invoice, image processing, booking system, and social network application products. We did not notice a product-related sensitivity around security tests. 

\textbf{Code Reviews (Peer Reviews)}
When asked about code/peer reviews, an important measure for security and privacy, nine participants said they have a process to review code written by the developers. One person said they regularly do it but not every time. The remaining six said they do not have a code review process, citing various reasons including lack of need or time. 

\textbf{Third-Party Integration}
All of the companies used third-party libraries in their products. When asked about the factors considered by the team when integrating third-party libraries, four participants said they check the popularity or the star count of the library. One reason was the easiness of finding solutions to problems when the library is widely-used. There is also less potential for future problems which was an important factor for two other participants regardless of the popularity of the software. Three of the participants mentioned documentation, maintenance, and general support provided by the library owners. Compatibility and speed were mentioned by three participants, two participants considered usability, and one participant mentioned good user interface. Two participants preferred open-source projects, since they can be adjusted. Other factors were ease of finding alternatives, price, and license of the software. One participant said that they use a third-party library for security. Only two participants mentioned security as a factor they consider when choosing a library. Two of the participants explicitly mentioned they do not consider security during the integration process. 

\subsection{Important Factors for Development}

To understand the importance of usability, security, and privacy as well as competing priorities, we asked our participants what is important to them when developing software. One prominent theme was solving a problem that could benefit a lot of people and being used by many. \textit{``First of all, it must be solving a problem. So there's no point in developing a software that doesn't solve a problem...After that, the fact that it will be used by people makes us a little happier'' (P11)}. One participant said the software should deliver what it promised.

Some participants answered our question with user-facing factors such as user experience and stability. \textit{``We try to maximize the user experience in terms of design'' (P12)}. A participant mentioned scalability as a factor while another mentioned fast performance.

One participant first cited the speed of development as the most important factor while developing. When asked to elaborate on other factors, they mentioned not having any technical debt and being bug-free. Afterwards, they said these factors were more important than speed. However, according to them speed of development still has higher priority compared to testing, citing the ever changing requests: \textit{``In the scenario where speed is important, testing is of little importance because the test you write today becomes irrelevant tomorrow when requests change. Speed, as I said, is more important than testing'' (P1)}.  

Other answers were focused on the quality and maintainability of the code. Writing reusable modular code was one of the most cited factors, as well as having proper documentation. As P13 put it, \textit{``The first, I think, is code quality and maintainability, code that the next developers can understand. Second, how is the documentation set up? I'm talking about the local environment. How to make a deployment? Unfortunately, these are not included in the projects. How is deployment done? How to test on a live server? Such documentation is also missing as far as I can see''}. One participant who is working on devices with limited memory said the size of the software should be small. 

Generally, we note that functional factors were mentioned, including some focus on usability, and that, except for service availability, participants rarely mentioned factors usually linked with security or privacy. We also did not notice any product-related sensitivity around mentioning security and privacy as important factors for development. 

\subsection{Consideration of Usability, Security, and Privacy} 

In an effort to find out how factors that are usually of secondary concern, but crucial to safe use of software, are handled and prioritized, we asked our participants how usability, security, and privacy are integrated into their products. The interview guide did not include security or privacy questions before this part of the interview. Eleven of the participants mentioned security mainly in contexts of third-party integration and testing. Only one participant mentioned privacy explicitly but five of them mentioned KVKK regarding the user data management before the questions around security and privacy were asked.

\subsubsection{Usability}
As usability of a system is a requirement for its secure and private use~\cite{garfinkel2014usable}, we were interested in processes that aim towards usability. While participants talked about implementing usability, none drew the connection to security or privacy. 
We saw various degrees to which usability was integrated into the software and development process. Receiving user feedback and having users in the loop was one of the distinct answers given by our participants. One participant mentioned having a modular code base that can be customized for each user and adapted to their requests quickly. Some of them mentioned thinking like a user while developing: \textit{``How can I use it more comfortably? We were personally looking at the project as end users. How can we use it more easily? I was also able to give feedback from the inside'' (P13)}.

Eight participants said that they have a person or a team responsible for the usability such as designers, front-end developers, or user experience teams. Some of these teams are outsourced, some are used only in the beginning of the development, some collaborate regularly during development. Notably, one participant considered usability a non-issue since their service is B2B.

One participant mentioned a developer who is curious and sensitive about accessibility who paid extra attention to create a framework that is accessible from the start. One participant mentioned that usability is something that may well not be developed in the early startup, but that can bring a company to rewrite their product as the company (and product) matures.

\subsubsection{Security}
For security, we received varying answers, with the dominant theme that functionality may be prioritized over security: Some (2) adhered to standards and walked us through best practices. One participant said security was central to their company, and mentioned a strong security mentality. Five of them talked about trusted (or untrusted) external components, like database management and cloud services. One participant said they received advice from experts and another mentioned outsourcing their security tests.

Access control was one of the prominent themes about security integration mentioned by four participants. Two cited encryption while some did not consider security in their development at all. One specifically said it was the users' responsibility. \textit{``Well, since we set it up in the customer environment, it is somewhat the responsibility of the customer, because firewall is customer's own firewall'' (P7)}. One participant mentioned that regulation mandated security implementations, while another spoke about the protection of system via developer security.

One participant (P11) put these varying degrees of caring about security in context: they explained that \textit{``there is no such thing as security in the first phase of the start-up,''} and that security is added at a later stage, as the company matures and customer or regulatory demands kick in. They said unless the startup's core product is related to security, security is probably a low priority: \textit{``This is how security usually proceeds in start-ups. I wrote the data and the code perfectly from the start. It is extremely safe. That is not believable anyway. So if a start-up tells you this, don't believe it. It is either a security start-up, or they are lying.''}.

\subsubsection{Privacy}
For privacy, three participants mentioned compliance with regulations such as KVKK. \textit{``One of the first things our CEO, says at such meetings (with customers) is: ``I talked to the lawyer before I talked to the developers.''. Compliance with KVKK and GDPR is a must for our work'' (P1)}. P1 also mentioned keeping their devices secure and encrypted to prevent access to their code.

One of the participants mentioned getting informed consent from users while another talked about the users' right to hide data from the service providers. Access control was again popular, mentioned by four participants, with one of the participants even mentioning keeping logs of access to data for possible audits. Similarly with the security, encryption of data (3) and trusted external services (2) was mentioned in privacy too.

Three of the participants called privacy a non-issue and one said they do not know the process, while one specifically said they had not thought about privacy \textit{yet}.
Another participant again summarized this as a maturity issue: that earlier in their startup, they did not have resources or awareness of privacy, but as the company matured, started to integrate it. \textit{``Honestly, we didn’t at first, due to our inexperience on the subject. We started to include it gradually'' (P6)}.

\subsection{Definition of Usability, Security, and Privacy} 
To better understand how participants conceptualize usability, security, and privacy, and therefore also, how they think about developing towards these goals, we asked our participants to define usability, security, and privacy in their own terms. Security was mainly defined as preventing unauthorized access, data protection, and being trustworthy. Privacy was also mainly defined with prevention of unauthorized access, as well as not sharing data with third parties, and data minimization. Our participants defined usability as ease of use, the system behaving in an expected way, and fulfilling its purpose.

\subsubsection{Usability}

One of the most mentioned concepts for usability was ease of use and access to functionality cited by six participants. \textit{``Ease of access in general. For example you need to access anything you want with a maximum of 3 clicks within the system. I think that three is good, four is normal, five is troublesome'' (P12).} Six of our participants also think a system that fulfills its purpose is usable. One participant said \textit{``Usability is actually about meeting a need. The easiest way to meet that need is usability for me'' (P11)} while three mentioned speed when asked to define usability. \textit{``I guess I would define it as making it possible for the user to reach what he wants to do in the fastest and shortest way possible'' (P1)}. A system behaving as expected is another factor given by three of our participants. \textit{``For usability: I'm actually looking for the save button at the bottom, that save button should be at the bottom. If you put the save button in the top right, it is not a usable application for me. I do scroll down, fill out the form below, then search for save button above. You know, everything needs to be in their right place. People have usage habits. This should be done according to the habit'' (P10)}. One person said that all three factors make a system usable.

Stability (2) is also important for our participants. \textit{``I think an application that works stably, and an application that is suitable for its purpose and that works stably is a usable application'' (P9)}. One person cited being inclusive as what makes a system usable. \textit{`` I can say that it is an application that should have a flow that does not impair the user experience of the usability of a product for both disabled and non-disabled people, without disrupting any user experience'' (P5)}. Taking feedback from users, not overwhelming them, and clearly communicating the functionality were mentioned by one participant. Thinking like a user (2) and engaging the user through notifications (1) was also given as the definition of a usable system.\textit{``We send a notification, if the user does open the application after seeing the content in the notification, then we have made a usable product'' (P3)}.

\subsubsection{Security}

One of the main themes our participants talked about was access control and prevention of unauthorized access to the code, servers, data, and so on. Six participants cited access control. \textit{``I define it as the inability of authorized persons-, more precisely, any person to perform any unauthorized action...The fact that no one but me can access the servers, would also be a branch of security... authorizations and roles, these are also security issues'' (P1)}. One participant cited not sharing the data with third parties, not changing user data without prior notice, and continuing the service as expected for users.

One participant explicitly mentioned the closeness of security and privacy. This was also reflected by participants talking about data protection (6) when talking about security. \textit{``When I think of security, I actually think of protecting someone’s data as committed. At the same time, I think it is related to the company that collect the data, and how that company invests in protecting that data, both on the basis of employees and external attacks.'' (P3)}. Another person also raised the issue of data leaking. \textit{``There is no such thing. Your data is in principle leaked by very large companies. At the moment, I know my name, surname, phone number, address, even my eating and drinking habits, what I bought last time, from whom I bought it, are on sale directly in the market. First of all, we cannot talk about security in such an environment'' (P11)}. 

Trust (4) is another factor cited by our participants. One participant correlated trust with widespread use of the system. \textit{``The things that I trust here are very few I can say, from applications. I don't know much about how data is processed. I would probably call the safe app the overused apps. But that wouldn't be the right answer either'' (P15)}. Another one mentioned absence of worry with trust. \textit{``It's got to be something that people can lay their back on and not have to worry about it'' (P8)}. Another mentioned the usage of trusted services for cloud and storage and that \textit{``the security they provide is actually our security'' (P14)}. 

Other mentions were fair distribution of tasks to developers and technical measures such as closing unused ports, using firewalls, and security-related routers. One participant talked about human factors while trying to ensure security. \textit{``Because security relies on the human factor, after all. For example, if I write down my username and password on a post-it and leave it on my desk, no matter how many software-related technical security measures I take, it doesn’t matter anymore'' (P6)}.

\subsubsection{Privacy}

As with security, prevention of unauthorized access was a prominent theme to describe privacy cited by seven of our participants. \textit{``Knowing the personal data I give is properly stored and not seen by people who are not supposed to see it'' (P3)}. Differently from when participants were discussing security definition, three raised the issue of third party access for privacy. \textit{``It is a matter of whether the data you share or hand over is transferred to the third party'' (P5)}. One participant mentioned not processing data without the consent of the user and another said not using the user data against them. \textit{``I think the basis of privacy is to ensure that something that can be used against a person is not used in that manner'' (P14)}.

Security was also brought up by six of our participants while defining privacy. \textit{``Privacy and security are essentially the same thing. They are not the same thing, but they look the same'' (P10)}. One participant distinguished privacy from security around how the data is handled. \textit{``Speaking as a user, if security is where my data is kept, privacy is how it is processed'' (P15)}.

Two participants described the concept of data minimization or need-to-know collection of data. \textit{``Knowing the personal data I give is properly stored and not seen by people who are not supposed to see it. Maybe instead of showing the whole thing for people who are supposed to see it-, You know how it happens in banks. The workers only see the, for an example, fourth letter of my mother's maiden name, and they don't know or see the whole thing. I think this is a good example of privacy'' (P3)}.

Two participants explicitly mentioned the KVKK (Personal Data Protection Authority). One participant, who is both the co-founder and the CTO of their company, mentioned reading through the KVKK to understand their responsibilities as a data controller. \textit{``I sat down and read the KVKK as the data controller, there were about 100 PDFs on KVKK’s website, I do not remember the address at the moment. I downloaded them one by one and read them while I was on the road, waiting for the bus, et cetera. I read them one by one to see if there is an adverse situation that prevents me from doing my job, or if there is anything that could put me in trouble as a data controller'' (P1)}. 

One participant defined privacy as information that is sensitive to the user and commented that they try to think like a user to decide on their improvements. \textit{``Would it be a problem for me if someone else had access this information of mine? If you say yes, it is true for the user as well. And we make our improvements by paying attention to them accordingly. I usually check it with myself. Would I feel bad? Yes, I would. Well, then that's not supposed to happen. It's a natural process, actually'' (P2)}.

\subsection{Problem \& Risks} 

Not all participants make strong distinctions between security and privacy. For example, when asked about security, they would talk about data leaking, and when discussing privacy, they would talk about attackers stealing data. One made the connection explicitly: \textit{``As we said, both security and privacy are close concepts'' (P9)}.

\subsubsection{Usability}
Many participants (7) mentioned (informal) user feedback; none mentioned formal usability testing. They improved their product's usability through customer feedback on usability issues. One mentioned \textit{``But our most important resource is the feedback of the end users'' (P12)}, another said \textit{``But when you put a light on the button where where the end user was supposed to press. So that literally decreased our technical support down to like 10 percent almost. So we definitely do run into errors of usability, but we try to resolve it as soon as possible with our team'' (P8)}. 

For usability, some (3) participants considered it a non-issue. This could be because they were early in the development process, had not yet developed a minimum viable product, and did not have access to users and user feedback yet. This is in line with the informal feedback collection from existing users employed in six other companies. Only two mentioned building usability in by design, e.g., \textit{``Since I am a backend programmer, I tried to shape these kinds of things by meeting with different companies twice'' (P1)}.

Some (2) talked about the challenges introduced by various skill and experience levels, and preferences by different types of users. \textit{``The biggest example or factor would be the computer usage habits of the user. Since not everybody can use computers effectively, we keep developing by focusing on: ``How can it be the best?''. For example, some of our customers use computers really well, and can easily use tabs to fill out the forms and move to the next tab. Some are so distant from computers that they may not even know how to do this, which is totally normal. How can we do it better, how can we do it easier, how can we create an easier flow? We constantly work on improvements. In other words, we need to consider every level when it comes to users'' (P6)}.

There was discussion of generational differences: \textit{``It is one of the most common problems users have lately. We are used to computers being used mostly with a keyboard and a mouse, but especially those who are born to the internet, which we call the Z generation, and those who use tablets and phones a lot, tend to use by swiping at the computer. It is one of the most common code problems we encounter'' (P11)}.

Some (2) mentioned backward compatibility / continuity in the UX, so that users would not be confused. \textit{`` The software, usability of the users. Maybe that's the thing, now when versions or new updates come, the user should not be alienated too much with the old version, you know, it shouldn't change too much. Even the front end you know, needs to be changed accordingly'' (P4)}. Supporting multiple devices is mentioned by one of the participants. No connections between usability and security/privacy were discussed.

\subsubsection{Security}
Security was discussed mostly among three dimensions: that having user data is a risk that invites data leaks, that malicious attackers could harm the company, and who should be responsible for security.

Collecting user data is seen as a security risk by six of our participants. \textit{`` this actually poses a security risk since we actually recorded this video. It's saved somewhere, after all. If it is not recorded, it is okay if it flows directly, but the fact that it is recorded poses a direct security risk for users'' (P9)}. User data leaks were mostly mentioned as an external threat that might impact users and the company. \textit{``Data leaking is the biggest problem for us. For example, a malicious person infiltrates the system and extracts that data. Even though it may not be very useful to them, but after all, since this is personal or commercial data, the issue of security is important'' (P6)}. Two participants mentioned introducing security gaps by integrating third-party libraries.

Attackers were mentioned by 10 participants, where two mentioned them as risks in situations where user data is not involved. For example, a participant who works with industrial robots stated there could be fatal outcomes of malicious attacks. \textit{``Any outside intervention in the system and access to the robot's codes pose a risk to the system. It poses risks for the entire factory, for the users.''} when asked in detail they said \textit{``Assuming you are carrying something that weighs three tons, unfortunately, a problem with the navigation system, the robot being displaced, or its sensor broken can result in fatal accidents'' (P15)}.

One participant felt that their B2B partner was responsible for security: \textit{``They have to do something. So they need to keep their servers more proper, more secure. That's why it doesn't concern us, quite frankly'' (P10)}. One said they as a company were responsible: \textit{``So let me tell you the simplest thing: the user is making a confidentiality agreement with you. You are actually responsible for the companies you trust and give data to. So it's your responsibility to keep track of it.''} but also that users share the risk: \textit{`` In fact, users share the risks we take. So the user can say: “you develop a software, and I trust you, but because of someone's mistake in your company, people know my data. This is not acceptable to me”. This is actually the user's risk'' (P11)}.

\subsubsection{Privacy}
Like security, privacy was discussed among the lines of attackers, vulnerabilities, and responsibility. Here, the question of consent and ethics was also discussed. For example, P6 said \textit{``The whole thing here is commercial data, and leakage of it is neither ethical nor desirable''} and P2 said \textit{`` This is because the user's data has been changed and updated without their consent''}.

Generally, privacy was more discussed as a prevention issue, namely as preventing leaks by preventing vulnerabilities (8), and two participants mentioned the possibility of attackers. Only two participants mentioned practices that could be understood as privacy-by-design or data minimization. \textit{``You may think that this is our data, we have flowing data, we only keep the username and password in our own database. In the rest of the system, the video is not saved anywhere unless the user requests it, and there is a structure that flows directly on the network traffic'' (P9)}.

When talking about responsibility, P3 mentioned dependency on external parties. \textit{``Depending on how well the product is protected against external attacks, data can be stolen. It depends a little bit on the services we use''}.

Participants also discussed preventing users from seeing other users' data, which may be based on our own example to prompt our users when they were stuck.

\subsection{Responsibility for Risks}

After asking our participants about usability, security, and privacy risks that could arise with regards to their products, we asked who is responsible for the risks they stated. Most of the participants (8) thought the responsibility is shared between multiple parties including developers, project leaders, companies, and users. A participant said  \textit{``I think the first is the responsibility of the software developers, the second is the responsibility of the leader and the people responsible for the software, and the third, I think, may not be included, but it is also the responsibility of the users'' (P13)}.

Two of the participants thought it was shared between developers and the companies where one of them stated that the users were also responsible. A participant, who is a co-founder of a company, said the responsibility is only with the company. On the other hand, two participants thought it was solely the responsibility of the developers and/or testers while another participant who is a CTO of the company thought the CTO is the sole responsible person. \textit{``I think the CTO is responsible for it as soon as anything that is not originally accessible from the outside can be accessed somehow'' (P1)}. 

Four participants mentioned users as a stakeholder in their answers, while one of them thought the users were the only responsible party, since users could turn off permissions or stop using the app completely. \textit{``I think it's the user's responsibility to reduce the risks...you will either not use it or you will not give these permissions'' (P9)}. Three participants thought the business to which they are providing their service was responsible. One of them, whose company provides smart glass applications to track factory workers' activity, stated that the white-collar workers such as the IT department (of the factory) were responsible for the risks. \textit{`` The confidentiality is the responsibility of the decision makers in the use of this work. Because how much privacy should be, that is, how much personal privacy should be? This is decided by the users of this application. I mean, the blue collar doesn't decide that'' (P10)}.

Aside from all these parties, a participant stated that there should be a suprastate organization that should provide guidelines and regulate them for everyone. \textit{`` I'm a little more strict with it. So there has to be a regulation. In other words, states, maybe even suprastate organizations, like World Health Organization that promise to reduce this risk in the pandemic, should be formed with the participation of all countries in the world'' (P11).} According to this participant, if there were a uniform guideline, even the startups who might have to use cheaper options would be protected.

\subsection{Steps to Mitigate Risks}
We asked our participants about what steps developers and companies take or should take to mitigate the risks participants discussed. 

\subsubsection{By Developers}
Two of our participants think developers should start the project with security and privacy in mind, P9 gave an example about the application permissions: \textit{``If the person who developed the application did not act aggressively and asked for permission only while using the application, for example, this could be the part of the event that looks at the software developer''}. P11 thought that developers should be aware of the implications of their actions : \textit{``You have to look at it from that perspective. The code you write touches someone's life somewhere.''}. Another theme raised by participants was that developers should improve themselves, keep themselves up to date with current technologies and security measures, learn security standards such as OWASP, and get acquainted with known security vulnerabilities. Participants mentioned that developers should know the sector they are providing service to and understanding the structure of the code they are working with.

One participant said that they try to avoid using third party applications and libraries in the code since it might be a possible risk to the system. If they must use external dependencies, they will need to keep up to date with the changes. Another one said developers should check these libraries for safety before using them. One participant mentioned their preference for using licensed software and tools.

There were also mentions of testing against possible attacks, keeping the computer and the code safe, and updating systems regularly. IP-based restrictions for access to data were cited. Participants mentioned that sensitive data such as user data should be stored encrypted and developers should not use logging with open sensitive data. One of the co-founder participants also said that the company should be able to lead the developers against these risks and the developers should follow directives of the company.

\subsubsection{By Companies}

One of the prominent themes our participants mentioned is having a system to check security of their products. Participants mentioned that this can be established by having a security team. Participants suggested that companies should work with a security firm or hire a consultant if they cannot afford to have a dedicated security team. They also made the suggestion that, if the company has limited resources, they could set up an open bug bounty program.  

Another theme was about giving sufficient funds to protect security, including providing tools the developers requested and giving security training to employees. The product should also be protected against malicious developers. As P15 explained, \textit{``As soon as there is a conflict of interest, the fate of the company is in the hands of the software developer. Maybe they can break the codes, get the codes and then use them themselves...there must be a layer of protection before an action is taken''}. 

Similarly, using licensed products and urging developers to keep up to date were mentioned as steps companies should take. Participants stated that companies also should give directives to their developers and manage the process. Other steps suggested by our participants are having a process to check the written code before merging, conducting security tests, carefully designing what information they should collect from users, keeping sensitive information safe, and deciding on the usage of third party apps carefully. One participant stated there were no risks mentioned while they worked there: \textit{``While I worked there, security risks were never mentioned. There was no such concern'' (P3)}.

\subsubsection{Special Role for Security \& Privacy}
Following the question about steps the companies should take, we also asked about whether there is a dedicated role who checks the software security and privacy in the company. 
For security, four participants said they do not have such a person in the company. Another five cited teamwork for these checks. For two of the companies, there is a dedicated person who checks the security of the software, but it is not in their job description. Two participants cited their CTO as the responsible party, while one referred to the co-founder of the company. Another participant stated that it is the responsibility of the senior developers. Lastly, there was a dedicated team responsible for security for one of the companies. 

For five companies, participants stated that there is no dedicated role responsible for privacy. For four other companies, participants again said that privacy is again teamwork (overlapping with those who said this for security). The previously mentioned CTOs responsible for security are also referred to for privacy. For three of the participants, we heard comments that there is such a person who unofficially took on the role. There is a dedicated team to check privacy of the product for two of the companies.

\subsubsection{Security \& Privacy Training}
We asked our participants whether they received any security and privacy training. Six of them received some kind of training either mandated by their company or of their own accord, four of them mentioned receiving online courses, while two of them completed university courses about S\&P.  One of the co-founders received counselling through a lawyer. One participant mentioned that they received a presentation from a security start-up. Three participants said that they never received any S\&P training.

We also asked if their company required them to receive S\&P training. Two of our participants said that their company requires training and pays for it. One participant said they regularly have self-improvement meetings with the whole team. The company did not require any training for the employees for the remaining 13, but four of these said they will provide training if there is a demand from the team.

\subsection{Turkey-specific Remarks by Participants}
Some participants commented specifically on how startups in Turkey function. We report these as they were stated, and recognize that these statements are informed by participants' lived experiences in Turkey, and while they were specifically made about Turkey, they may well be common to software development, or development in startups in general. We highlight that experiences with local laws and data leaks may influence development practices in a way that may impact products that reach beyond the local context.
While discussing the code reviews, one of the participants mentioned how there was no apparent hierarchy in the startups of Turkey. The co-founder or CTO often work alongside with the developers to solve the problems the team faces while writing the code. The distribution of responsibility was defined as ``horizontal''.

Data leaks were commonly discussed with concepts related to privacy and security. While acknowledging that this is a problem worldwide, even for big platforms, participants also commented on the data leaks happening specifically in Turkey. Some of the participants were wary of the data leaks as founders and developers since the ``sanctions are heavy'' (P11) and the data leakage is ``something that's taken very seriously in Turkey, according to law'' (P8). On the other hand, the mistrust against the companies as consumers were mentioned. P11 remarked that even their eating and drinking habits are leaked and being sold. There were also some comparisons between KVKK and GDPR around data leaks. While the sanctions are regarded as heavy, the process to enforce data protection happens after there is a problem. P11 mentioned how KVKK was dealing with the aftermaths of a data leak while GDPR ensures that the protections are in place before the business is approved for operation.

\section{Discussion}
In our interviews with 16 Turkish startup software developers, security and privacy seemed desirable but secondary to the primary goal of having a viable product. They were implemented later in the process of building the startup. We saw that more mature startups did consider security and privacy, and had dedicated teams or persons for them, or outsourced them to third parties. We find that, except for considering both KVKK and GDPR, Turkish startup developers' security and privacy perceptions and processes resemble those found in published literature based on interviews with developers from various company sizes in Germany, the United States of America, the United Kingdom, Brazil, Israel, and China~\cite{gutfleisch2022usable, xie2011programmers,sheth2014us,souza2017software,hadar2018privacy,zhang2018toward}.

In their interview study with developers from different size companies, Gutfleisch et al.~\cite{gutfleisch2022usable} found that usability was considered as an important aspect of developing but it is not considered in the context of security and privacy. We also find that usability is important for startup developers in our study. It is also easier to define usability concepts for the developers in our study compared to privacy and security. Similarly with Hadar et al.~\cite{hadar2018privacy}, we find that developers equate privacy with data security. When prompted, our participants usually listed security and privacy risks that could happen as a result of an outsider attack rather than an insider (e.g. an authorized employee with data access). There is a lack of privacy knowledge by developers~\cite{senarath2019will} which can be a result of the consideration of privacy as a non-functional requirement~\cite{senarath2018developers} and prioritizing learning about concepts around functional requirements before privacy.

\textbf{We find that in the early phases of the startup, with limited resources, a low number of employees that ``wear many hats'', and a focus on creating a functional product, it is not easy for companies to integrate security and privacy to their products.} Unless the main focus of the product is around security and privacy, these concepts may not be considered. Since prior work shows company sizes affect the privacy-levels~\cite{balebako2014privacy} and adoption of privacy preserving technologies~\cite{hadar2018privacy}, it is expected that startups might give less priority to security and privacy. They may gain priority when the minimum viable product (MVP) is released and there is a solid customer base, or government-mandated rules are enforced such as GDPR and KVKK. External pressures enable startup companies to implement measures to ensure security and protect privacy~\cite{assal2018security}. Some participants reported their companies were actively engaged with Turkish privacy laws, as well as EU privacy laws. This is not surprising, as most participants stated that their company was either providing services abroad or was planning to in the future.

\textbf{We also found that third-party libraries were frequently used but most of the companies did not have a clear guideline for the approval of these libraries.} As with Zhang et al.~\cite{zhang2018toward}, we found that possible issues around security and privacy are not considered while integrating third-party libraries as long as the desired functionality can be achieved. However, advertisements were not used widely by the companies since most of them were providing B2B services.

From our interviews, we noticed that spinouts (i.e. startups split from bigger companies) or startups founded by developers who worked in corporate companies had more grounded pipelines for development with the company organization and hierarchy well-defined. \textbf{Considering skills of the team drives the practices~\cite{tahaei2021privacy, souza2021survey}, carry-over from previous experiences prove to be valuable for development, as well as security and privacy practices.}

Most of the products in our study were focused on a B2B model. Future work can focus on the security and privacy perceptions of developers who provide services to end-users. Products facing end-users might have different features than B2B products. For example, none of the companies in our study used advertisements in their products. Products facing end-users might utilize advertisement as a business model where these ad networks use dark patterns to nudge developers to make less privacy-conscious choices~\cite{tahaei2021developers}.

\textbf{Based on our findings, we make recommendations for five groups of actors who can take action to improve security, privacy, and usability in the Turkish startup software development as well as startups in general}: industry, developers, researchers, educators, and regulators.

\subsection{Industry}

\textbf{Companies, including startups, should allocate sufficient funds from the start for usability, security, and privacy~\cite{yilmaz2020protecting} and implement the processes early on~\cite{cavoukian2009privacy}}. Starting a product with security and privacy in mind is a preventive practice that is a longtime investment, which might be one of the reasons why startups rarely do it. While startups have limited funds, foregoing security and privacy risks can have more serious implications in the future. 

Hiring developers that are experienced in security and privacy is another way to improve the product. Skill set and experiences of developers influence the mindset of their team and organization~\cite{haney2018we,tahaei2021privacy}. This is especially important in startups where the teams are small. Our participants also mentioned having senior developers in the team shaping the development pipeline from the start, bringing their expertise gained from other companies to the startup.

Companies should relay clearly that they value security and privacy. Bu et al.~\cite{bu2020privacy} shows that employment of best practices should be a top-to-down process that is properly incentivized. The companies should provide a clear roadmap to developers and give them necessary training to achieve the goals of the company regarding security and privacy. In addition to training, companies should also provide an environment for developers to implement the things they've learned~\cite{poller2017can}. Division of responsibility and roles should also be clear. Some of our participants were unclear who (within their company) was responsible for security and privacy, some were even unclear whether another company or their users were responsible. However, some of them had dedicated people, teams, or external contractors responsible for security and privacy.

Participants also mentioned that the libraries and components they use from outside their company should be secure and trustworthy. For industry in general, more transparency and clear communication about the security and privacy benefits of shared components (such as libraries) can save others time and effort for testing, and improve overall security and privacy. This may help with improving startup software security and privacy while still considering the scarce resources of startups. We concur that security along the software supply chain is a long-standing issue that is currently being addressed by an U.S. executive order~\cite{executiveorder}, as well as efforts by US companies~\cite{googleopensource}. Interestingly but not unexpectedly, these geographically remote efforts may have broad impacts into software produced in Turkey. 

\subsection{Developers}

\textbf{Many of our participants approach security and privacy issues not as design issues, but something that can happen \textit{to them}, for example through attacks by third-parties.} This is in line with findings of Hadar et al.~\cite{hadar2018privacy}, who found that their participants used the terms of data security for privacy which led them to limit privacy to third party attacks. However, when asked about responsibilities for these risks, most of our participants saw themselves among the responsible parties. This finding tracks with Xiao et al.~\cite{xiao2014social} who found that participants felt responsible for S\&P and at odds with findings of Xie et al.~\cite{xie2011programmers} and Bednar et al.~\cite{bednar2019engineering}, whose participants reject responsibility.

Prior work has shown that developers are influenced by their teammates~\cite{tahaei2021privacy}. For example, they adopt tools they learn from their peers~\cite{xiao2014social}, and having a ``Champion'' in the team can drive the team and company to be more security and privacy conscious~\cite{haney2019motivating,tahaei2021privacy}. Developers embed their own values in the products they design and implement~\cite{shilton2018engaging} and they have an important role for compliance with local and global regulations~\cite{yilmaz2020protecting}.

\subsection{Researchers}

Most of the research conducted in the usable security and privacy domain is West-centric~\cite{wang2017third} whereas a substantial amount of developers do not study, live, and work in these countries~\cite{SOdevelopersurvey21}. There is a need for more inclusive research~\cite{wang2018inclusive,henrich2010most}. International standards which can shape the actions of the companies should also be taken into consideration. Our participants mentioned compliance with GDPR~\cite{gdpr} as well as KVKK~\cite{kvkk}. While existing research with developers in non-EU countries mentions GDPR~\cite{hadar2018privacy}, there is minimal mention and research on security and privacy laws effective in other countries. As a research community that does not only focus on western contexts, we could learn from the implementation successes and failures of these laws (e.g. KVKK in Turkey, which was rolled out in 2016 before GDPR became enforceable in 2018), and use that knowledge to improve security and privacy globally.

\textbf{As researchers, we need to clearly communicate findings to make them available to global audiences.} Using community-sourced information like stack overflow might not be enough for developers in terms of security~\cite{acar2016you} and the official guidelines shared over the internet are often outdated and unhelpful~\cite{acar2017developers}. There is also the added barrier of language where the developer might not have access to guidance in their language. As we see from our participants, developers that do not necessarily speak English provide services worldwide. Hence, it is important for these developers to have guidance they could understand so they can implement state-of-the-art practices easily. As researchers from these countries, it is one of our responsibilities to help our communities to have access to these practices.

\subsection{Educators}

We see that only two of our participants completed a course related to security and privacy in the university. This may be due to the fact that many topics important for real-world software development remain advanced electives in many curricula. \textbf{We think that security, privacy, and usability should be introduced early in the curricula.} University courses and projects can inspire students to have the security mindset~\cite{schenier2008} which will enable them to influence their companies~\cite{tahaei2021privacy}. Most software developers we interviewed have Bachelor's degrees, so if these important classes are ``advanced'' electives, those who write software may not have the necessary education to implement best practices. This also includes structured secure software engineering, including having a software development lifecycle/pipeline. Having developers with Security \& Privacy education is especially important for startups, where the skill sets of core developers steer the direction of the product in the early stages. It is also important to teach concepts of usability, privacy, and security in non-CS classes, as many developers cross over from related or unrelated fields. Finally, this reworking of curricula must take place worldwide, as classes (not) taught in e.g., Turkey may influence the security of software products used globally. 

\subsection{Regulators}

As with the Assal and Chiasson~\cite{assal2018security}, we also find that external pressures from customers and regulations influence the adoption of security practices in the development process. Adherence to data protection laws (e.g. KVKK and GDPR) was promising, where participants mentioned compliance when discussing privacy. In order for privacy to not be ignored, participants mentioned that having clear laws to follow and adhere to make them prioritize privacy, at least to the point where they adhere with KVKK. One of our CTO participants said that they read through all of the KVKK documents, even though they also received counselling from lawyers, since they are responsible as the data controller. Analogously to privacy, we can imagine that similar laws and ``minimal standards'' for security may make a difference for secure development. We think that \textbf{it is important to make these regulations \textit{accessible} to developers}: in some instances, participants mentioned that they wanted to comply with KVKK but were struggling due to the volume of documents and readability, and one mentioned that they had to involve a lawyer. Especially for startups this overhead can be punitive or prohibitive~\cite{norval2021data}.

\section{Conclusion}
We conducted interviews with 16 developers who work in Turkish software startups to understand the state of awareness, risk assessment, and development practices related to security and privacy in Turkish startups. We find that developers struggle with defining security and privacy, may not have dedicated software development processes, responsibilities, or training to address those. 
Implementing secure development practices and protecting user privacy should be adopted widely despite the perception that security and privacy is less important to startups than a working product.
We think that this is only possible by change enacted by multiple actors: prioritizing security and privacy in regulation, education, and budgeting, as well as making them easier and feasible for individual developers to implement. Research that takes the global developer population into account may be able to  evaluate the effects of locally realized guidelines more quickly, which can feed back into better security and privacy globally.

\section*{Acknowledgments}
We would like to thank Lucy Simko, Anna Lena Rotthaler, and Yoshi Kohno for their support of this work, and MPI-SP for supporting this research in many ways, including employment and a remote internship.

\bibliographystyle{plain}

\bibliography{references}

\begin{thebibliography}{10}

\bibitem{acar2016you}
Yasemin Acar, Michael Backes, Sascha Fahl, Doowon Kim, Michelle~L Mazurek, and
  Christian Stransky.
\newblock You get where you're looking for: The impact of information sources
  on code security.
\newblock In {\em 2016 IEEE Symposium on Security and Privacy (SP)}, pages
  289--305. IEEE, 2016.

\bibitem{acar2017developers}
Yasemin Acar, Christian Stransky, Dominik Wermke, Charles Weir, Michelle~L
  Mazurek, and Sascha Fahl.
\newblock Developers need support, too: A survey of security advice for
  software developers.
\newblock In {\em 2017 IEEE Cybersecurity Development (SecDev)}, pages 22--26.
  IEEE, 2017.

\bibitem{alhazmi2020struggle}
Abdulrahman Alhazmi and Nalin Asanka~Gamagedara Arachchilage.
\newblock Why are developers struggling to put {GDPR} into practice when
  developing {Privacy-Preserving} software systems?
\newblock USENIX Association, August 2020.

\bibitem{assal2018security}
Hala Assal and Sonia Chiasson.
\newblock Security in the software development lifecycle.
\newblock In {\em Fourteenth Symposium on Usable Privacy and Security
  ($\{$SOUPS$\}$ 2018)}, pages 281--296, 2018.

\bibitem{assal2019think}
Hala Assal and Sonia Chiasson.
\newblock 'think secure from the beginning' a survey with software developers.
\newblock In {\em Proceedings of the 2019 CHI conference on human factors in
  computing systems}, pages 1--13, 2019.

\bibitem{balebako2014privacy}
Rebecca Balebako, Abigail Marsh, Jialiu Lin, Jason~I Hong, and Lorrie~Faith
  Cranor.
\newblock The privacy and security behaviors of smartphone app developers.
\newblock 2014.

\bibitem{bednar2019engineering}
Kathrin Bednar, Sarah Spiekermann, and Marc Langheinrich.
\newblock Engineering privacy by design: Are engineers ready to live up to the
  challenge?
\newblock {\em The Information Society}, 35(3):122--142, 2019.

\bibitem{startupblink}
Startup Blink.
\newblock Global startup ecosystem index 2022, 2022.

\bibitem{bozkurt2017bilicsim}
Ba{\c{s}}ak Bozkurt and Aylin AKPINAR.
\newblock Bili{\c{s}}im sekt{\"o}r{\"u}nde toplumsal cinsiyete dayali i{\c{s}}
  b{\"o}l{\"u}m{\"u}.
\newblock {\em Marmara {\"U}niversitesi Kad{\i}n ve Toplumsal Cinsiyet
  Ara{\c{s}}t{\i}rmalar{\i} Dergisi}, 1(2):17--28, 2017.

\bibitem{brennan1981coefficient}
Robert~L Brennan and Dale~J Prediger.
\newblock Coefficient kappa: Some uses, misuses, and alternatives.
\newblock {\em Educational and psychological measurement}, 41(3):687--699,
  1981.

\bibitem{bu2020privacy}
Fei Bu, Nengmin Wang, Bin Jiang, and Huigang Liang.
\newblock “privacy by design” implementation: Information system
  engineers’ perspective.
\newblock {\em International Journal of Information Management}, 53:102124,
  2020.

\bibitem{cavoukian2009privacy}
Ann Cavoukian.
\newblock Privacy by design.
\newblock 2009.

\bibitem{corbin2014basics}
Juliet Corbin and Anselm Strauss.
\newblock {\em Basics of qualitative research: Techniques and procedures for
  developing grounded theory}.
\newblock Sage publications, 2014.

\bibitem{donaldson2002understanding}
Stewart~I Donaldson and Elisa~J Grant-Vallone.
\newblock Understanding self-report bias in organizational behavior research.
\newblock {\em Journal of business and Psychology}, 17(2):245--260, 2002.

\bibitem{garfinkel2014usable}
Simson Garfinkel and Heather~Richter Lipford.
\newblock Usable security: History, themes, and challenges.
\newblock {\em Synthesis Lectures on Information Security, Privacy, and Trust},
  5(2):1--124, 2014.

\bibitem{garousi2015survey}
Vahid Garousi, Ahmet Co{\c{s}}kun{\c{c}}ay, Aysu Betin-Can, and Onur
  Demir{\"o}rs.
\newblock A survey of software engineering practices in turkey.
\newblock {\em Journal of Systems and Software}, 108:148--177, 2015.

\bibitem{womendevelopers}
Ecem~Korkmaz Gelal.
\newblock Women in turkish software development report, 2021.

\bibitem{giardino2014we}
Carmine Giardino, Michael Unterkalmsteiner, Nicolo Paternoster, Tony Gorschek,
  and Pekka Abrahamsson.
\newblock What do we know about software development in startups?
\newblock {\em IEEE software}, 31(5):28--32, 2014.

\bibitem{octoverse}
GitHub.
\newblock Empowering healthy communities, 2021.

\bibitem{googleopensource}
Google.
\newblock Introducing secure open source pilot, 2021.

\bibitem{green2016developers}
Matthew Green and Matthew Smith.
\newblock Developers are not the enemy!: The need for usable security apis.
\newblock {\em IEEE Security \& Privacy}, 14(5):40--46, 2016.

\bibitem{gutfleisch2022usable}
Marco Gutfleisch, Jan~H. Klemmer, Niklas Busch, Yasemin Acar, M.~Angela Sasse,
  and Sascha Fahl.
\newblock {How Does Usable Security (Not) End Up in Software Products? Results
  From a Qualitative Interview Study}.
\newblock In {\em Proc.\ 43rd IEEE Symposium on Security and Privacy (SP'22)}.
  IEEE, 2022.

\bibitem{hadar2018privacy}
Irit Hadar, Tomer Hasson, Oshrat Ayalon, Eran Toch, Michael Birnhack, Sofia
  Sherman, and Arod Balissa.
\newblock Privacy by designers: software developers’ privacy mindset.
\newblock {\em Empirical Software Engineering}, 23(1):259--289, 2018.

\bibitem{haney2019motivating}
Julie~M Haney and Wayne~G Lutters.
\newblock Motivating cybersecurity advocates: Implications for recruitment and
  retention.
\newblock In {\em Proceedings of the 2019 on Computers and People Research
  Conference}, pages 109--117, 2019.

\bibitem{haney2018we}
Julie~M Haney, Mary Theofanos, Yasemin Acar, and Sandra~Spickard Prettyman.
\newblock " we make it a big deal in the company": Security mindsets in
  organizations that develop cryptographic products.
\newblock In {\em Fourteenth Symposium on Usable Privacy and Security
  ($\{$SOUPS$\}$ 2018)}, pages 357--373, 2018.

\bibitem{henrich2010most}
Joseph Henrich, Steven~J Heine, and Ara Norenzayan.
\newblock Most people are not weird.
\newblock {\em Nature}, 466(7302):29--29, 2010.

\bibitem{howard2006security}
Michael Howard and Steve Lipner.
\newblock {\em The security development lifecycle}, volume~8.
\newblock Microsoft Press Redmond, 2006.

\bibitem{executiveorder}
Joseph R.~Biden Jr.
\newblock Executive order on america's supply chains, 2021.

\bibitem{klein2018many}
Richard~A Klein, Michelangelo Vianello, Fred Hasselman, Byron~G Adams,
  Reginald~B Adams~Jr, Sinan Alper, Mark Aveyard, Jordan~R Axt, Mayowa~T
  Babalola, {\v{S}}t{\v{e}}p{\'a}n Bahn{\'\i}k, et~al.
\newblock Many labs 2: Investigating variation in replicability across samples
  and settings.
\newblock {\em Advances in Methods and Practices in Psychological Science},
  1(4):443--490, 2018.

\bibitem{kvkk}
Ki\c{s}isel Verileri~Koruma Kurumu.
\newblock Ki\c{s}isel verilerin korunmas\i{} kanunu, 2016.

\bibitem{lazar2017research}
Jonathan Lazar, Jinjuan~Heidi Feng, and Harry Hochheiser.
\newblock {\em {Research Methods in Human-Computer Interaction}}.
\newblock Morgan Kaufmann, 2017.

\bibitem{liang2021embracing}
Calvin~A Liang, Sean~A Munson, and Julie~A Kientz.
\newblock Embracing four tensions in human-computer interaction research with
  marginalized people.
\newblock {\em ACM Transactions on Computer-Human Interaction (TOCHI)},
  28(2):1--47, 2021.

\bibitem{mcdonald2019reliability}
Nora McDonald, Sarita Schoenebeck, and Andrea Forte.
\newblock Reliability and inter-rater reliability in qualitative research:
  Norms and guidelines for cscw and hci practice.
\newblock {\em Proceedings of the ACM on human-computer interaction},
  3(CSCW):1--23, 2019.

\bibitem{muthukrishna2020beyond}
Michael Muthukrishna, Adrian~V Bell, Joseph Henrich, Cameron~M Curtin,
  Alexander Gedranovich, Jason McInerney, and Braden Thue.
\newblock Beyond western, educated, industrial, rich, and democratic (weird)
  psychology: Measuring and mapping scales of cultural and psychological
  distance.
\newblock {\em Psychological science}, 31(6):678--701, 2020.

\bibitem{norval2021data}
Chris Norval, Heleen Janssen, Jennifer Cobbe, and Jatinder Singh.
\newblock Data protection and tech startups: The need for attention, support,
  and scrutiny.
\newblock {\em Policy \& Internet}, 2021.

\bibitem{ccpa}
State of~California.
\newblock California consumer privacy act (ccpa), 2018.

\bibitem{SOdevelopersurvey21}
Stack Overflow.
\newblock Stack overflow developer survey 2021, 2021.

\bibitem{paternoster2014software}
Nicol{\`o} Paternoster, Carmine Giardino, Michael Unterkalmsteiner, Tony
  Gorschek, and Pekka Abrahamsson.
\newblock Software development in startup companies: A systematic mapping
  study.
\newblock {\em Information and Software Technology}, 56(10):1200--1218, 2014.

\bibitem{poller2017can}
Andreas Poller, Laura Kocksch, Sven T{\"u}rpe, Felix~Anand Epp, and Katharina
  Kinder-Kurlanda.
\newblock Can security become a routine? a study of organizational change in an
  agile software development group.
\newblock In {\em Proceedings of the 2017 ACM conference on computer supported
  cooperative work and social computing}, pages 2489--2503, 2017.

\bibitem{rader2020have}
Emilee Rader, Samantha Hautea, and Anjali Munasinghe.
\newblock " i have a narrow thought process": Constraints on explanations
  connecting inferences and self-perceptions.
\newblock In {\em Sixteenth Symposium on Usable Privacy and Security
  ($\{$SOUPS$\}$ 2020)}, pages 457--488, 2020.

\bibitem{saunders2018saturation}
Benjamin Saunders, Julius Sim, Tom Kingstone, Shula Baker, Jackie Waterfield,
  Bernadette Bartlam, Heather Burroughs, and Clare Jinks.
\newblock Saturation in qualitative research: exploring its conceptualization
  and operationalization.
\newblock {\em Quality \& quantity}, 52(4):1893--1907, 2018.

\bibitem{schlesinger2017intersectional}
Ari Schlesinger, W~Keith Edwards, and Rebecca~E Grinter.
\newblock Intersectional hci: Engaging identity through gender, race, and
  class.
\newblock In {\em Proceedings of the 2017 CHI conference on human factors in
  computing systems}, pages 5412--5427, 2017.

\bibitem{schenier2008}
Bruce Schneier.
\newblock The security mindset, 2008.

\bibitem{senarath2018developers}
Awanthika Senarath and Nalin~AG Arachchilage.
\newblock Why developers cannot embed privacy into software systems? an
  empirical investigation.
\newblock In {\em Proceedings of the 22nd International Conference on
  Evaluation and Assessment in Software Engineering 2018}, pages 211--216,
  2018.

\bibitem{senarath2019will}
Awanthika Senarath, Marthie Grobler, and Nalin Asanka~Gamagedara Arachchilage.
\newblock Will they use it or not? investigating software developers’
  intention to follow privacy engineering methodologies.
\newblock {\em ACM Transactions on Privacy and Security (TOPS)}, 22(4):1--30,
  2019.

\bibitem{sheth2014us}
Swapneel Sheth, Gail Kaiser, and Walid Maalej.
\newblock Us and them: a study of privacy requirements across north america,
  asia, and europe.
\newblock In {\em Proceedings of the 36th International Conference on Software
  Engineering}, pages 859--870, 2014.

\bibitem{shilton2018engaging}
Katie Shilton.
\newblock Engaging values despite neutrality: Challenges and approaches to
  values reflection during the design of internet infrastructure.
\newblock {\em Science, Technology, \& Human Values}, 43(2):247--269, 2018.

\bibitem{maxqda}
VERBI Software.
\newblock Maxqda 2022, 2021.

\bibitem{souza2021survey}
Renata Souza, Orges Cico, and Ivan Machado.
\newblock A survey on software engineering practices in brazilian startups.
\newblock {\em arXiv preprint arXiv:2108.00343}, 2021.

\bibitem{souza2017software}
Renata Souza, Karla Malta, and Eduardo~Santana De~Almeida.
\newblock Software engineering in startups: a single embedded case study.
\newblock In {\em 2017 IEEE/ACM 1st International Workshop on Software
  Engineering for Startups (SoftStart)}, pages 17--23. IEEE, 2017.

\bibitem{eurostat}
Eurostat Statistics.
\newblock Ict specialists in employment, 2021.

\bibitem{tahaei2021privacy}
Mohammad Tahaei, Alisa Frik, and Kami Vaniea.
\newblock Privacy champions in software teams: Understanding their motivations,
  strategies, and challenges.
\newblock In {\em Proceedings of the 2021 CHI Conference on Human Factors in
  Computing Systems}, pages 1--15, 2021.

\bibitem{tahaei2019survey}
Mohammad Tahaei and Kami Vaniea.
\newblock A survey on developer-centred security.
\newblock In {\em 2019 IEEE European Symposium on Security and Privacy
  Workshops (EuroS\&PW)}, pages 129--138. IEEE, 2019.

\bibitem{tahaei2021developers}
Mohammad Tahaei and Kami Vaniea.
\newblock “developers are responsible”: What ad networks tell developers
  about privacy.
\newblock In {\em Extended Abstracts of the 2021 CHI Conference on Human
  Factors in Computing Systems}, pages 1--11, 2021.

\bibitem{gdpr}
The~European Union.
\newblock General data protection regulation, 2016.

\bibitem{wang2017third}
Yang Wang.
\newblock The third wave? inclusive privacy and security.
\newblock In {\em Proceedings of the 2017 New Security Paradigms Workshop},
  pages 122--130, 2017.

\bibitem{wang2018inclusive}
Yang Wang.
\newblock Inclusive security and privacy.
\newblock {\em IEEE Security \& Privacy}, 16(4):82--87, 2018.

\bibitem{xiao2014social}
Shundan Xiao, Jim Witschey, and Emerson Murphy-Hill.
\newblock Social influences on secure development tool adoption: why security
  tools spread.
\newblock In {\em Proceedings of the 17th ACM conference on Computer supported
  cooperative work \& social computing}, pages 1095--1106, 2014.

\bibitem{xie2011programmers}
Jing Xie, Heather~Richter Lipford, and Bill Chu.
\newblock Why do programmers make security errors?
\newblock In {\em 2011 IEEE symposium on visual languages and human-centric
  computing (VL/HCC)}, pages 161--164. IEEE, 2011.

\bibitem{yilmaz2019observed}
Murat Y{\i}lmaz.
\newblock Observed effects of software processes change in three software
  firms: industrial exploratory case study.
\newblock 2019.

\bibitem{yilmaz2020protecting}
Tolga Y{\i}lmaz, S{\"u}leyman~Muhammed Ar{\i}kan, Fatma Su, and {\"O}zg{\"u}r
  Y{\"u}rekten.
\newblock Protecting personal information in enterprise applications.
\newblock In {\em 2020 Turkish National Software Engineering Symposium (UYMS)},
  pages 1--4. IEEE, 2020.

\bibitem{zhang2018toward}
Rui Zhang and Genying Xie.
\newblock Toward understanding iot developers in chinese startups.
\newblock In {\em Proceedings of the 22nd international conference on
  evaluation and assessment in software engineering 2018}, pages 181--186,
  2018.

\end{thebibliography}

\appendices

\section{Survey Questions}
\label{app:prescreen}
Participants were shown a participant information sheet in Turkish explaining the survey and the following interview if they consented to the study. After getting the consent of the participants, they were directed to the prescreen.
\subsection{Prescreen}
\begin{enumerate}
    \item What country are you a citizen of? (this is weird wording in English)
    \begin{itemize}
        \item List of Countries
    \end{itemize}
    \item (If Turkey is not selected on Q1) Have you been living in Turkey more than 10 years?
    \begin{itemize}
        \item Yes
        \item No (Screen out the participants if selected)
    \end{itemize}
    \item What is the location of the company you are currently working at?
    \begin{itemize}
        \item List of Countries (Screen out the participants if Turkey is not selected)
    \end{itemize}
    \item How old is the company you are currently working at?
    \begin{itemize}
        \item Number Entry (Screen out the participants if the number is higher than 5)
    \end{itemize}
\end{enumerate}
\subsection{Company Information and Demographics}
\begin{enumerate}
    \item Can you tell us a bit about your experience on working in startups? (2-3 sentences)
    \item What is your role in the company you are currently working at?
    \item What is the sector of the company you are currently working at?
    \item How many employees are there in the company?
    \begin{itemize}
        \item 1-10
        \item 11-50
        \item 51-100
        \item 101-500
        \item 501-1000
        \item 1000+
    \end{itemize}
    \item How many years have you been working as software developer? (Please answer using digits, e.g. 3)
    \item How many members are in your team other than you? (Please answer using digits, e.g. 3)
    \item What is the highest degree or level of school you have completed? (If you’re currently enrolled in school, please indicate the highest degree you have received.)
    \begin{itemize}
        \item High school graduate or less
        \item Some college or two-year associate degree
        \item Bachelor’s degree 
        \item Master’s degree
        \item PhD degree
        \item Professional Degree (e.g. doctor of medicine)
        \item Prefer to not disclose
        \item I never completed any formal education
    \end{itemize}
    \item If you indicated that you received a degree, what was the field of study?
    \item How old are you? (Please answer using digits, e.g. 3)
    \item What is your gender?
    \begin{itemize}
        \item Woman
        \item Man
        \item Non-binary
        \item Prefer not to disclose
        \item Prefer to self-describe
    \end{itemize}
    \item Which language would you prefer for the interview? 
    \begin{itemize}
        \item English
        \item Turkish
    \end{itemize}
\end{enumerate}

\section{Interview Guide}
\label{app:guide}
This is the English version of the interview guide.

\subsection{BACKGROUND AND EDUCATION}
\begin{enumerate}
    \item You say you are working as a $<$job role from screener$>$. Can you tell us a bit about your background? 
    \begin{itemize}
        \item How did you come to work in the field and how long have you been working?
        \item How did you learn software development? 
        \item Have you received formal education? 
    \end{itemize}
    \item What resources are you using to learn new concepts?
\end{enumerate}

\subsection{CURRENT POSITION AND THE PROJECT}
\begin{enumerate}
    \item Tell us a bit about your current project? 
    \begin{itemize}
        \item Who are the users?
        \item What kind of development are you doing? 
        \item In which programming language?
    \end{itemize}
    \item You said your team had XX people in the survey. How do you divide the responsibilities?
    \begin{itemize}
        \item What is your role in the project?
    \end{itemize}
\end{enumerate}

\subsection{DEVELOPMENT LIFECYCLE}
\begin{enumerate}
    \item What is the typical pipeline for developing software (in the company)? 
    \begin{itemize}
        \item Can you walk us through the process? 
        \item What tools are you using to develop the product?
        \item What tools are you using to manage teamwork? Are you doing peer reviews?
        \item What is the procedure of integrating third-party libraries?
    \end{itemize}
    \item Who set up this development process? 
    \begin{itemize}
        \item How did they decide on it?
    \end{itemize}
    \item What resources are you using to look up when you face problems in the code?
    \item What is important to you when you are developing a piece of software? 
    \item (What) do you test your software for?
    \item Do you collect any user data and how are they stored if so? Who has access to it? 
\end{enumerate}

\subsection{SECURITY AND PRIVACY}
\begin{enumerate}
    \item How is usability integrated into the development process? Is it even considered?
    
    \begin{itemize*}
        \item And security?
        \item And privacy?
    \end{itemize*}
    \item How would you define usability? 
    
    \begin{itemize*}
        \item And security?
        \item And privacy?
    \end{itemize*}
    \item Do you run into usability problems when you develop software? 
    \begin{itemize}
        \item How about your users?
    \end{itemize}
    \item What are some security risks you might face in your job? 
    \begin{itemize}
        \item For your software? 
        \item What are some of the risks users might face using your service?
    \end{itemize}
    \item And privacy risks? 
    
     \begin{itemize*}
        \item For your software? 
        \item For the users? 
    \end{itemize*}
    \item Who do you think is responsible for mitigating such risks?
    \item What steps the developers are taking to mitigate security/privacy risks?
    \item What steps the company is taking to mitigate security/privacy risks?
    \item Is there someone responsible for making sure the software is secure?
    \begin{itemize}
        \item Can you walk me through how they do it?
    \end{itemize}
    \item And for privacy?
    \item Have you taken any education/training in software security and privacy or used learning resources around security and privacy? 
    \begin{itemize}
        \item If yes, what were those?
        \item Does the company require you to take security/privacy courses?
    \end{itemize}
    \item Are you using Ads (Advertisements) in your products? (If yes) Are you aware of the security and privacy policies of those services?
    \item Should we have asked any other questions about how you do secure development?
\end{enumerate}

\section{Codebook}
\label{app:codebook}

\begin{itemize}
    \item Company
    \begin{itemize}
        \item Product
        \item Users of the Product
        
        \begin{itemize*}
            \item End-User
            \item B2B
            \item Location
        \end{itemize*}
    \end{itemize}
    \item Development Pipeline
    \begin{itemize}
            \item Development Process
            
            \begin{itemize*}
                \item Very Organized
                \item Some Organization
                \item No Organization
            \end{itemize*}
            \item Builder of the Pipeline
            
            \begin{itemize*}
                \item Founders
                \item Teamwork
                \item Defined Person
                \item No Pipeline
            \end{itemize*}
            \item Code Reviews
            \item Third-Party Integration
    \end{itemize}
    \item Testing
    \begin{itemize}
            \item In-house Testing Team
            \item Developers as Testers
            \item Another Person in Company Testing
            \item Outsourced Testing Team
            \item Outsider Person or Customer Testing
            \item Security Testing
            \item Functional Testing
    \end{itemize}
    \item Important Factors Developing
    \item Integration into Software
    
    \begin{itemize*}
            \item Usability
            \item Security
            \item Privacy
    \end{itemize*}
    \item Definition of Concepts
    
    \begin{itemize*}
            \item Usability
            \item Security
            \item Privacy
    \end{itemize*}
\item Problems \& Risks 

    \begin{itemize*}
            \item Usability
            \item Security
            \item Privacy
    \end{itemize*}
    \item Responsibility for Risks
    \begin{itemize}
            \item Special Role - Security
            \item Special Role - Privacy
    \end{itemize}
    \item Steps to Mitigate
    \begin{itemize}
            \item Developers
            \item Companies
    \end{itemize}
    \item Security \& Privacy Education 
    \begin{itemize}
        \item Required by Company
    \end{itemize}
\end{itemize}

\end{document}